\documentstyle [12pt,ere96] {article}
\def\R{{\rm I \!\!\, R}}
\def\be{\begin{equation}}
\def\ee{\end{equation}}
\def\bea{\begin{eqnarray}}
\def\eea{\end{eqnarray}}

\begin{document}
\heading{PERFECT FLUID SPACE-TIMES ADMITTING A 3-DIMENSIONAL
CONFORMAL GROUP ACTING ON NULL ORBITS}

\author{A. M. Sintes   and  J. Carot} 
{Departament de F\'{\i}sica, Universitat Illes Balears, 07071 Palma de
Mallorca   Spain }

\begin{abstract}
\baselineskip .34cm
Space-times admitting a 3-dimensional Lie group of conformal motions $C_3$
acting on null orbits are studied. Coordinate expressions for the metric and
the conformal Killing vectors (CKV) are then provided (irrespectively of the
matter content) and all possible perfect fluid solutions are found, although
none of them verifies the weak and dominant energy conditions over the whole
space-time manifold. \end{abstract}

\section{Introduction}
So far in the literature,
 the study of null orbits has been restricted to isometries
only. The groups $G_r$, $r\ge 4$, on $N_3$ have at least one subgroup $G_3$
which may act on $N_3$, $N_2$ or $S_2$ \cite{Kramer}. For $G_3$ on $S_2$,
one obtains special cases of the LRS models, $G_r$
 admitting either a group $G_3$
on $N_3$ or a null Killing vector \cite{Barnes73}. The case $G_3$ on $N_2$
was studied by Barnes \cite{Barnes79}, the group $G_3$ is then of Bianchi type
$II$ and perfect fluid solutions are excluded since the metric 
leads to a Ricci tensor whose Segre type is not that of a perfect fluid.
Another case that has been considered in the literature is that
of a $G_3$ on $N_3$, such case is subject to the condition $R_{ab}k^ak^b=0$
and this condition excludes perfect fluids with $\mu+p\not=0$. Perfect
fluid solutions cannot admit a non-twisting $(w=0)$ null Killing vector
except if $\mu+p=0$. The algebraically special perfect fluid solutions with
twisting null Killing vectors are treated by Wainwright \cite{Wain70} and
 they
admit an Abelian group $G_2$.

This paper will deal with space-times admitting a 3-dimensional Lie group of
conformal motions $C_3$ acting on null orbits.
In the
 beginning one could get the feeling that this kind of space-times would be
empty for perfect fluid solutions, since the line element of these space-times
is, by the theorem of Defrise-Carter \cite{Defrise}, conformally related to one
admitting a $G_3$ acting on null orbits and these ones, as we have pointed out
above, do not admit perfect fluid solutions. But, as we will show, this is
not the case, since indeed a conformal scaling does  change the Ricci
tensor, but there are just a few solutions.

\section{Space-times admitting CKVs acting 
on null  orbits}

We shall concern ourselves with
space-times $(M,g)$
that admit a three-parameter conformal group $C_3$ containing an Abelian
two-parameter subgroup of isometries $G_2$, whose orbits $S_2$ are spacelike,
diffeomorphic to $\R^2$ and admit orthogonal two-surfaces;
furthermore, we shall assume that the $C_3$  acts transitively on null
orbits $N_3$.

The classification of all possible Lie algebra structures for  ${\cal C}_3$
under the previous hypothesis was given in \cite{Ali3} where
coordinates were adapted so that the line element  associated with the metric 
$g$
can be written as \cite{Wain}  
\be
ds^2=e^{2F}\{-dt^2+dr^2+Q[H^{-1}(dy+Wdz)^2+Hdz^2]\}\ ,
\label{n6}
\ee
where $F$, $Q$, $H$ and $W$ are all functions of $t$ and $r$ alone.

If the conformal 
algebra ${\cal C}_3$ belongs to the family (A), it was shown in \cite{Ali3}
 that, for null conformal orbits, one can always bring $X$ to the form 
\be
 X=\partial_t+\partial_r+X^y(y,z)\partial_y+X^z(y,z)\partial_z \ , \label{n2}
\ee
where $X^y(y,z)$ and $X^z(y,z)$ are linear functions of their arguments to be
determined from the commutation relations of $X$ with the Killing vectors.
Specializing now the conformal equations
to the CKV (\ref{n2}) and the metric (\ref{n6}),  
 for each possible
case, one has the following forms for $X$ and the metric
functions $F$, $Q$, $H$, and $W$ appearing in (\ref{n6})
\bea
(I)&\,& Q=q(t-r), \quad H=h(t-r), \quad W= w(t-r), \nonumber\\
&\,& X=\partial_t+\partial_r. \\
(II)&\,& Q=q(t-r), \quad H=h(t-r), \quad W= w(t-r)-{t+r \over 2}, \nonumber\\
&\,& X=\partial_t+\partial_r+z\partial_y.\\
(III)&\,& Q=e^{-{t+r \over 2}}q(t-r), \quad H=e^{{t+r \over 2}} h(t-r), \quad
W=e^{{t+r \over 2}} w(t-r), \nonumber\\ &\,& X=\partial_t+\partial_r
+y\partial_y. \label{n13}\\
 (IV)&\,& Q=e^{-(t+r)}q(t-r), \quad H=h(t-r), \quad W=
w(t-r)-{t+r \over 2}, \nonumber\\ &\,& X=\partial_t+\partial_r
+(y+z)\partial_y+z\partial_z.\\
 (V)&\,& Q=e^{-(t+r)}q(t-r), \quad
H=h(t-r), \quad W= w(t-r), \nonumber\\ &\,&
X=\partial_t+\partial_r +y\partial_y+z\partial_z. \\ 
(VI)&\,&
Q=e^{-(1+p){t+r \over 2}}q(t-r), \quad H=e^{(1-p){t+r \over 2}}h(t-r),  \quad
W=e^{(1-p){t+r \over 2}} w(t-r), \nonumber\\ &\,& X=\partial_t+\partial_r
+y\partial_y+pz\partial_z\quad (p\not=0,1).\label{n16} \\ 
(VII)&\,& Q=e^{-p{t+r \over 2}}q(t-r),\quad 
c=c(t-r), \quad g=g(t-r), \nonumber\\
 &\,& H={{\sqrt{4-p^2}\over 2}\over \sqrt{1+c^2+g^2} 
+c\cos(\sqrt{4-p^2}{t+r \over 2}) +g\sin(\sqrt{4-p^2}{t+r \over 2})} , 
\nonumber\\ &\,& W={p\over 2}+{{\sqrt{4-p^2}\over 2}
[c\sin(\sqrt{4-p^2}{t+r \over 2})-g\cos(\sqrt{4-p^2}{t+r \over 2})] \over
\sqrt{1+c^2+g^2} +c\cos(\sqrt{4-p^2}{t+r \over 2}) +g\sin(\sqrt{4-p^2} {t+r
\over 2})}, \nonumber\\ &\,& X=\partial_t+\partial_r
-z\partial_y+(y+pz)\partial_z\quad (p^2<4). \label{n18}\eea
In all of these cases $F=F(t,r)$ and the conformal factor $\Psi$ is given by
\be
\Psi=F_{,t}+F_{,r} \   .
\ee

Furthermore one can prove that family (B) cannot admit conformal
Killing vectors acting on null orbits (the proof can be found  in
\cite{tesis}).

Note that these results are completely independent of
the Einstein field equations and therefore of
the assumed energy-momentum tensor. 

\section{Perfect fluid solutions}
For perfect fluid solutions the study is exhausted. For a maximal $C_3$, with
a proper CKV, all possible solutions have been found (see \cite{tesis} for
details).  They correspond only to the types $III$ and $VI$, although none of
them satisfies the weak and dominant energy conditions over the whole
space-time manifold.

{\bf Type $VI$} (this includes the type $III$ for $p=0$)  

We make the coordinate transformation  $u=t+r$ and $v=t-r$, so that we have
$h=h(v)$ and $q=q(v)$. The field equations yield
\be
W=0 \ ,
\ee
\be
F=f(x)+{1\over 2}{1+p \over 1-p}\ln h -{1\over 2} \ln q \ ,\quad x\equiv
u-{2\over 1-p}\ln h \ , \label{n60}
\ee
\be
0=\left\{ {q_{,v}h_{,v} \over qh}+{h_{,vv}\over h}\right\}\Sigma_0
+\left({h_{,v}\over h}\right)^2\Sigma_1\ ,\label{n61}
\ee
where
\bea
\Sigma_0&\equiv&-1+p^4+4f_{,x}-4pf_{,x}+4p^2f_{,x}-4p^3f_{,x}+8f_{,x}^2
-8p^2f_{,x}^2 \nonumber\\
 & -&32f_{,x}^3+32pf_{,x}^3-8f_{,xx}+8p^2f_{,xx}+32f_{,xx}f_{,x}
-32pf_{,xx}f_{,x}
\eea
\bea
\Sigma_1&\equiv &2+2p+2p^2+2p^3-16f_{,x}-8pf_{,x}-16p^2f_{,x}-8p^3f_{,x}
+32f_{,x}^2 +16pf_{,x}^2 \nonumber\\
&+&48p^2f_{,x}^2-64pf_{,x}^3-16f_{,xx}+16pf_{,xx}-32pf_{,xx}+64pf_{,xx}f_{,x}\ .
\eea
 $h_{,v}=0$ is excluded since the solution
 does not correspond to a perfect fluid. Therefore, two possibilities arise:
\bea
&{\rm i})& \quad \Sigma_0=0, \quad \Sigma_1=0 \nonumber \\
&{\rm ii})&\quad {\displaystyle q_{,v}h_{,v} \over\displaystyle qh}+
{\displaystyle h_{,vv}\over\displaystyle h}=a \left({\displaystyle
h_{,v}\over\displaystyle h}\right)^2 \qquad (a={\rm const})\ . \nonumber 
\eea
 In the first case $f_{,x}$ must be a constant, and
therefore the CKV is not proper. For the second case we have
\be
{q_{,v}\over q}=a{h_{,v}\over h}-{h_{,vv}\over h_{,v}}\ , \label{n64}
\ee
which can be integrated to give
\be
q={h^a \over h_{,v}}\ , \label{n65}
\ee
and equation (\ref{n61}) reduces to:
\be
1={f_{,xx}[f_{,x}32(ap-a-2p)+8(2-p^2a-2p+4p^2+a)]\over [4f_{,x}-p-1]
[f_{,x}^2 8(ap-a-2p)+f_{,x}8(p^2+1)+a-ap+ap^2-ap^3-2-2p^2]}\  .\label{n66} \ee
It is convenient to divide the analysis into three sub-cases. \hfill\break

\underline{Sub-case (a):} $a=2p/(p-1)$.

Equation (\ref{n66})  can be readily integrated to give
\be
f={p+1 \over 4}x- {(1-p)^2\over p^2+1}{1\over 2}\ln\vert x\vert +c\ , \quad
c={\rm const} \ . \label{n69}
\ee
We notice that for $p=-1$ there exists a third Killing vector of the form
\be
\zeta=\left( {1\over 2}+{1\over 2}{h\over h_{,v}}\right) \partial_t+
\left({1\over 2}-{1\over 2}{h\over
h_{,v}}\right) \partial_r+y\partial_y-z\partial_z \ .\label{n70}
\ee
 \hfill\break

\underline{Sub-case (b):} $a=2/(1-p$).
 
When $p=-1$ the solution is a particular case of  sub-case (a). The
remaining cases may  now be integrated giving:
\be
f=-\ln\vert 1-e^{-(1+p)x/4} \vert +c \ ,\quad
c={\rm const} \ .\label{n73}
\ee
We note that this sub-case admits the further Killing vector
\be
\zeta=\left( {1\over 2}+{1-p\over 4}{h\over h_{,v}}\right)\partial_t+ \left(
{1\over 2}-{1-p\over 4}{h\over h_{,v}}\right)\partial_r+ {1-p\over
2}y\partial_y -{1-p\over
2}z\partial_z \ , \label{n74}
\ee
which violates our requirement of a maximal three-dimensional
conformal  group $C_3$.
 \hfill\break

\underline{Sub-case (c)}: we finally consider the possibility
$a\not=2p/(p-1)$ and $a\not=2/(1-p)$.

The solution of (\ref{n66}) is then given implicitly by
\be
x=\gamma_1\ln\vert f_{,x}-\beta_0\vert + \gamma_2\ln\vert f_{,x}-\beta_+\vert +
\gamma_3\ln\vert f_{,x}-\beta_-\vert \label{n79}
\ee
where
\be
\beta_0={p+1\over 4}\ ,\quad
\beta_{\pm}={-2(p^2+1)\pm\sqrt{2(p^2+1)(1-p)^2(a^2-2a+2)}\over 4(ap-a-2p)}\ .
\label{n76}
\ee
 $\gamma_i$, $i=1,2,3$  being constants.

A careful analysis of the energy conditions shows that for all cases (i.e.,
for all values of the parameters
$a$ and $p$) the solutions can only satisfy the energy
conditions over certain open domains of the manifold (see \cite{tesis}).

 \acknowledgements{This work has been supported by DGICYT
Research Project No. PB94-1177.}

\vfill
\end{document}